\documentclass[aps,prl,twocolumn,showpacs,superscriptaddress,groupedaddress]{revtex4}  
\usepackage{graphicx}  
\usepackage{dcolumn}   
\usepackage{bm}        
\usepackage{amssymb}   
\usepackage{amsmath}

\begin{document}


\title{One-way surface magnetoplasmon cavity
and its application for nonreciprocal devices}

\author{Kexin Liu} \affiliation{Department of Electromagnetic Engineering, School of Electrical Engineering, Royal Institute of Technology KTH, Stockholm S-100 44, Sweden}
\affiliation{Centre for Optical and Electromagnetic Research,
Zhejiang Provincial Key Laboratory for Sensing Technologies, JORCEP
(Sino-Swedish Joint Research Center of Photonics), Zhejiang
University, Hangzhou 310058, China}

\author{Amir Torki} \affiliation{Department of Electromagnetic Engineering, School of Electrical Engineering, Royal Institute of Technology KTH, Stockholm S-100 44,
Sweden}

\author{Sailing He}
\thanks{correspongding author: sailing@kth.se} \affiliation{Department of Electromagnetic
Engineering, School of Electrical Engineering, Royal Institute of
Technology KTH, Stockholm S-100 44, Sweden} \affiliation{Centre for
Optical and Electromagnetic Research, Zhejiang Provincial Key
Laboratory for Sensing Technologies, JORCEP (Sino-Swedish Joint
Research Center of Photonics), Zhejiang University, Hangzhou 310058,
China}
\vskip 0.25cm

\begin{abstract}
We theoretically analyze surface magnetoplasmon modes in a compact
circular cavity made of magneto-optical material under a static
magnetic field. Such a cavity provides two different physical
mechanisms for the surface wave to circulate in a unidirectional
manner around the cavity, which offers more freedom to realize
one-way surface wave. We also show the interaction between this
one-way cavity and waveguides, through an example of a circulator,
which lays the fundamental groundwork for potential nonreciprocal
devices.
\end{abstract}

 \maketitle

Surface plasmons (SPs) at metal-dielectric interfaces enable
confining and manipulating electromagnetic waves on a subwavelength
scale \cite{1}. A variety of applications based on the concept of
SPs have been achieved at different frequency regimes such as
focusing of light beyond the diffraction limit \cite{2},
ultracompact nano-lasers \cite{3} and nano-antennas \cite{4,5}.
Applying a static magnetic field on SPs breaks the time-reversal
symmetry and gives rise to the nonreciprocal propagation of SPs
called surface magnetoplasmons (SMPs) \cite{6,7,8}. The asymptotic
frequencies of SMPs in the forward and backward directions are split
by the external magnetic field. This provides a one-way regime for
SMPs, i.e., the propagation of the surface wave is only allowed in
one direction within the frequency range between two asymptotic
frequencies \cite{9,10}. There are also other mechanisms supporting
one-way waves \cite{11,12,13,14,15,16} including topological
protected edge states in photonic crystal. Compared with these
however, the mechanism based on SMPs seems more attractive due to
its simple configuration.

Recent research has shown that SMPs are immune to backscattering
from disorder \cite{9,17}, which might be useful for applications
involving isolators. Most results are based on SMPs at a planar
interface and use only a single frequency range
\cite{9,10,17,18,19,20}. Since the cavity structure is always
compact and at the heart of many photonic components (e.g. High-Q
surface-plasmon-polariton whispering-gallery microcavity \cite{21}),
it is interesting to study the behavior of SMPs in cavity structures
and investigate the physical mechanisms for a one-way SMP cavity
mode, i.e., when SMPs circulate around the cavity in a unidiretional
manner. Integrating such SMP structures with conventional waveguides
is also of particular interest for constructing some nonreciprocal
devices.

In this letter, we theoretically analyze SMP modes in a compact
circular cavity made of magneto-optical (MO) material under a static
magnetic field and demonstrate two different physical mechanisms for
one-way cavity modes. The first one results from splitting the
asymptotic frequencies for clockwise and anti-clockwise cavity
modes, which forms the upper one-way frequency range. The other is
obtained by different cut-off frequencies for the lowest clockwise
and anti-clockwise cavity modes, which forms the lower one-way
frequency range. The circulation directions of SMPs in the cavity
are opposite in these two different mechanisms. We also study the
interaction between such a one-way SMP cavity and conventional
waveguides, which is fundamental for designing nonreciprocal
devices. A compact circulator is designed and shown by direct
numerical simulations as an example.

We firstly analyze the cavity modes for SMPs to get an insight into
the possible one-way regimes. The cavity structure is a circle with
radius $R$ surrounded by air in a two-dimensional (2D) system as
shown in the insert of Fig. 1(a). The circular cavity is made of a
semiconductor which is an MO material at THz frequencies. By
applying a static magnetic field in the +z direction and assuming
the material is lossless, the relative permittivity tensor of the
semiconductor takes the following form in cylindrical coordinates
$(\rho ,\varphi ,z)$ \cite{6}:
\begin{equation}
\varepsilon  = \left( {\begin{array}{*{20}{c}}
{{\varepsilon _1}}&{i{\varepsilon _2}}&0\\
{ - i{\varepsilon _2}}&{{\varepsilon _1}}&0\\
0&0&{{\varepsilon _3}}
\end{array}} \right)\
\end{equation}
with ${\varepsilon _1} = {\varepsilon _\infty }(1 - \frac{{\omega
_p^2}}{{{\omega ^2} - \omega _c^2}})$, ${\varepsilon _2} =
{\varepsilon _\infty }\frac{{{\omega _c}\omega _p^2}}{{\omega
({\omega ^2} - \omega _c^2)}}$, ${\varepsilon _3} = {\varepsilon
_\infty }(1 - \frac{{\omega _p^2}}{{{\omega ^2}}})$, where $\omega $
is the angular frequency, ${\omega _p}$ is the plasma frequency of
the semiconductor, ${\omega _c} = eB/{m^ * }$ is the electron
cyclotron frequency ($e$ and ${m^ * }$ are, respectively, the charge
and effective mass of the electron, and $B$ is the applied magnetic
field), and  ${\varepsilon _\infty }$ is the high-frequency
permittivity of the semiconductor. We only consider the TM mode
$({H_z},{E_\rho },{E_\varphi })$ in this cavity, where ${E_z} =
{H_\rho } = {H_\varphi } = 0$. Maxwell's equations yield
\begin{subequations}
\begin{equation}
{E_\varphi } = \frac{1}{{i\omega {\varepsilon _0}(\varepsilon _1^2 -
\varepsilon _2^2)}}(\frac{{i{\varepsilon _2}}}{\rho }\frac{\partial
}{{\partial \varphi }} - {\varepsilon _1}\frac{\partial }{{\partial
\rho }}){H_z}
\end{equation}
\begin{equation}
{E_\rho } = \frac{1}{{i\omega {\varepsilon _0}(\varepsilon _1^2 -
\varepsilon _2^2)}}(\frac{{{\varepsilon _1}}}{\rho }\frac{\partial
}{{\partial \varphi }} + i{\varepsilon _2}\frac{\partial }{{\partial
\rho }}){H_z}
\end{equation}
\begin{equation}
\frac{1}{\rho }\frac{\partial }{{\partial \rho }}\rho {E_\varphi } -
\frac{1}{\rho }\frac{\partial }{{\partial \varphi }}{E_\rho } =  -
i\omega {\mu _0}{H_z}
\end{equation}
\end{subequations}
Taking the field ${H_z} = \psi (\rho ){e^{im\varphi }}$, where $m$
is an integer indicating the azimuthal mode number, we obtain the
following differential equation from Eq. (2):
\begin{equation}
[\frac{\partial }{{\partial {\rho ^2}}} + \frac{1}{\rho
}\frac{\partial }{{\partial \rho }} + ({k^2} - \frac{{{m^2}}}{{{\rho
^2}}})]\psi (\rho ) = 0
\end{equation}
where ${k^2} = {\omega ^2}{\mu _0}{\varepsilon _0}{\varepsilon _v}$
, with Voigt permittivity ${\varepsilon _v} = {\varepsilon _1} -
{{\varepsilon _2^2} \mathord{\left/
 {\vphantom {{\varepsilon _2^2} {{\varepsilon _1}}}} \right.
 \kern-\nulldelimiterspace} {{\varepsilon _1}}}$. Inside the circle $(\rho  \le R)$, the
solution for Eq. (3) is the Bessel function of the first kind
${J_m}(k\rho )$. Outside the circle $(\rho  > R)$, for outgoing
waves, the solution for Eq. (3) is the Hankel function of the first
kind $H_m^{(1)}({k_0}\rho )$, where ${k_0}$ is the wave vector in
air. Therefore, the solution is given by
\begin{equation}
\psi (\rho ) = \left\{ {\begin{array}{*{20}{c}}
  {A \cdot {J_m}(k\rho )}\qquad \ \ \ {\rho  \leqslant R} \\
  {C \cdot H_m^{(1)}({k_0}\rho)\qquad {\rho  > R}      }
\end{array}} \right.
\end{equation}
According to the boundary conditions that ${H_z}$ and ${E_\varphi }$
are continuous at $\rho  = R$, we obtain the following
eigenfrequency equation of the cavity modes:
\begin{equation}
\frac{{{J_m}(kR)}}{{H_m^{(1)}({k_0}R)}} = \frac{{({\varepsilon
_2}m/R) \cdot {J_m}(kR) + {\varepsilon _1}k \cdot
{{J'}_m}(kR)}}{{(\varepsilon _1^2 - \varepsilon _2^2){k_0} \cdot
H{{_m^{(1)}}^\prime }({k_0}R)}}
\end{equation}
We can solve Eq. (5) numerically for each given azimuthal mode
number $m$ and obtain the dispersion relation between the
eigenfrequency and the mode number $m$. The linear term
$({\varepsilon _2}m/R) \cdot {J_m}(kR)$ with respect to $m$, which
originates from the off-diagonal element of the permittivity tensor,
breaks the left-right symmetry of the dispersion relation.
Therefore, the solutions for Eq. (5) and also the field
distributions are different for the cavity modes $\left| { + m}
\right\rangle $ and $\left| { - m} \right\rangle $, where $\left| {
+ m} \right\rangle $ and $\left| { - m} \right\rangle $ represent
the modes with positive and negative mode numbers, respectively. In
addition, we only investigate the fundamental radial mode, since it
is the most confined of the radial modes.

\begin{figure}
\includegraphics[scale=0.53]{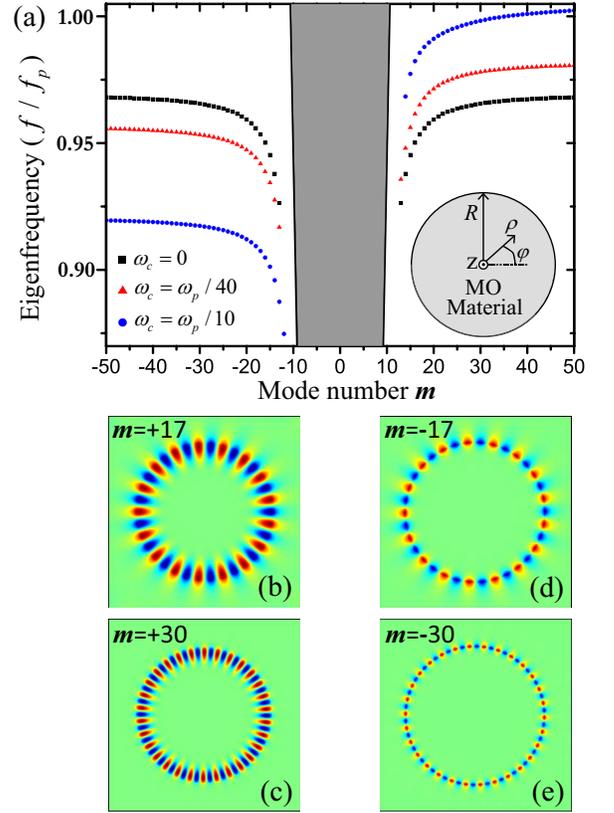}
\caption{\label{fig:epsart} (a) The dispersion relation between the
eigenfrequency and azimuthal mode number $m$ of the circular cavity.
The circular cavity with radius $R$ is surrounded by air in the
two-dimensional (2D) system as shown in the insert. The cavity is
made of the semiconductor InSb, which is a magneto-optical (MO)
material at THz frequencies. Three cases with increasing electron
cyclotron frequency ${\omega _c} = 0$ (black square dots), ${\omega
_c} = {\omega _p}/40$ (red triangular dots) and ${\omega _c} =
{\omega _p}/10$ (blue circular dots) are plotted. The gray region
represents the region above the light line. The distributions of the
modal field $H_z$ are shown in (b) $m=+17$, (c) $m=+30$, (d) $m=-17$
and (e) $m=-30$, respectively, when ${\omega _c} = {\omega _p}/10$.}
\end{figure}

We take a semiconductor disk made of InSb with $R=250$ $\mu$m as an
example to show the solution for Eq. (5). At room temperature, the
semiconductor InSb takes the following parameters: ${\varepsilon
_\infty } = 15.6$, ${f_p} = {\omega _p}/2\pi  = 2$ THz and
$m^*=0.014m_0$ ($m_0$ is the free electron mass) \cite{22}. We study
three cases with increasing applied magnetic field  $B = 0$ T
(${\omega _c} = 0$), $B = 0.025$ T (${\omega _c} = {\omega _p}/40$)
and  $B = 0.1$ T (${\omega _c} = {\omega _p}/10$) along the $ + z$
axis and solve Eq. (5) numerically. The solutions for those three
cases are illustrated in Fig. 1(a), where the dots show the
dispersion relations and the gray region represents the region above
the air light line. The air light line is determined by $f = {{mc}
\mathord{\left/ {\vphantom {{mc} {2\pi R}}} \right.
 \kern-\nulldelimiterspace} {2\pi R}}$. The dispersion
relations start from different cutoff frequencies and approach
different asymptotic frequencies in these three cases. Firstly, in
the absence of an external magnetic field ($B = 0$ T, ${\omega _c} =
0$, black square dots), the dispersion relation is symmetric, as the
time-reversal symmetry is preserved. When $m \to \pm \infty $, the
mode has an infinite azimuthal wave vector, and the asymptotic
frequency is ${f_s} = {{({\omega _p}} \mathord{\left/
 {\vphantom {{({\omega _p}} 2}} \right.
 \kern-\nulldelimiterspace} 2}\pi ) \sqrt {{{{\varepsilon _\infty }} \mathord{\left/
 {\vphantom {{{\varepsilon _\infty }} {({\varepsilon _\infty } + {\varepsilon _{air}})}}} \right.
 \kern-\nulldelimiterspace} {({\varepsilon _\infty } + {\varepsilon _{air}})}}} $, which is the same result for SPs
without magnetization. When $m$ is very small, due to the limitation
of the air light line, the corresponding eigenfrequency can be a
complex number, which corresponds to a radiative mode \cite{23}. The
lowest mode with a real eigenfrequency gives the cutoff frequency
for nonradiative modes. In this letter, we only consider the
nonradiative modes. Note that when $R \to \infty $, the cutoff
frequency will approach zero, which is exactly the case for SPs at a
planar interface. The curvature of the interface enables one to
achieve the cutoff frequency above zero. For the second case with $B
= 0.025$T (${\omega _c} = {\omega _p}/40$, red triangular dots), the
dispersion relation is asymmetric, and the asymptotic frequencies
are split by the external magnetic field. The $ + m$ branch of the
dispersion relation is higher than the dispersion relation with $B =
0$ T, while the $ - m$ branch is lower than the one with $B = 0$ T.
When $m \to  \pm \infty $, applying the formulas ${J_{m \to {\rm{ +
}}\infty }}(s) \sim ({1 \mathord{\left/
 {\vphantom {1 {\sqrt {2\pi m} }}} \right.
 \kern-\nulldelimiterspace} {\sqrt {2\pi m} }}) \cdot {(es/2m)^m}$,
$H_{m \to {\rm{ + }}\infty }^{(1)}(s) \sim ( - i\sqrt 2 /\sqrt {\pi
m} ) \cdot {(es/2m)^{ - m}}$, ${J_{ - m}}(s) = {( - 1)^m}{J_m}(s)$
and ${H_{ - m}}(s) = {( - 1)^m}{H_m}(s)$ in Eq. (5), we obtain the
following asymptotic frequencies

\begin{equation}
{f_{ \pm \infty }} = \frac{1}{{4\pi }}(\sqrt {\omega _c^2 + 4\omega
_p^2\frac{{{\varepsilon _\infty }}}{{{\varepsilon _\infty } +
{\varepsilon _{air}}}}} ) \pm {\omega _c})
\end{equation}
which are the same as SMPs at a planar interface discussed in
\cite{17}. Within the one-way frequency range ${f_{ - \infty }} < f
< {f_{ + \infty }}$, there only exists $\left| { + m} \right\rangle
$ modes, and SMPs can only circulate around the cavity
anti-clockwise. Furthermore, there does exist different cutoff
frequencies for $\left| { + m} \right\rangle $ and $\left| { - m}
\right\rangle $ cavity modes. The cutoff frequency ${f_{ + c}}$ of
the $\left| { + m} \right\rangle $ mode is higher than the cutoff
frequency ${f_{ - c}}$ of the $\left| { - m} \right\rangle $ mode,
so there is another one-way frequency range. Within ${f_{ - c}} < f
< {f_{ + c}}$, there only exists $\left| { - m} \right\rangle $
modes, and SMPs can only circulate around the cavity clockwise. In
the third case, we apply a stronger magnetic field $B = 0.1$ T
(${\omega _c} = {\omega _p}/10$, blue circular dots). The applied
strong magnetic field considerably breaks the symmetry of the
dispersion relation. The cutoff frequency ${f_{ + c}}$ of the
$\left| { + m} \right\rangle $ mode can be lifted higher than the
asymptotic frequency ${f_{ - \infty }}$ of the $\left| { - m}
\right\rangle $ mode. In this case, the modes $\left| { + m}
\right\rangle $ and $\left| { - m} \right\rangle $ are separated in
completely different frequency ranges. The $\left| { + m}
\right\rangle $ modes are within ${f_{ + c}} < f < {f_{ + \infty }}$
and SMPs can only circulate anti-clockwise in this frequency range.
The $\left| { - m} \right\rangle $ modes are within ${f_{ - c}} < f
< {f_{ - \infty }}$ and SMPs can only circulate clockwise in this
frequency range. The circulation directions are opposite in these
two different one-way frequency ranges. This interesting feature
offers more freedom, both in the one-way direction and in the
frequency range, for designing nonreciprocal components.

To verify the theoretical results obtained from Eq. (5), we also
solve the eigenfrequency of the cavity mode and calculate the modal
field distributions by a finite element method (FEM) in the
commercial software COMSOL. The deviation of the FEM results from
the results obtained by Eq. (5) is less than
$1{\raise0.5ex\hbox{$\scriptstyle 0$} \kern-0.1em/\kern-0.15em
\lower0.25ex\hbox{$\scriptstyle {00}$}}$. The modal field (${H_z}$)
distributions of the cavity modes $\left| { \pm 17} \right\rangle $
and $\left| { \pm 30} \right\rangle $ for $B = 0.1$ T are plotted in
Fig. 1(b)-1(d), respectively. The field is confined at the boundary
of the cavity and this confinement increases with $\left| m
\right|$. We also note that the field distributions are different
for the $\left| { + m} \right\rangle $ and $\left| { - m}
\right\rangle $ modes, since the strong external magnetic field
breaks the symmetry. To clearly demonstrate the circulation
direction of SMPs, we made videos of the harmonic wave propagation
around the cavity at the eigenfrequencies of modes $\left| { \pm 17}
\right\rangle $. To excite these modes, a magnetic current point
source is placed at the boundary of the cavity. The field ${H_z}$ of
the wave is recorded (see supplementary material online).

\begin{figure}
\includegraphics[scale=0.4]{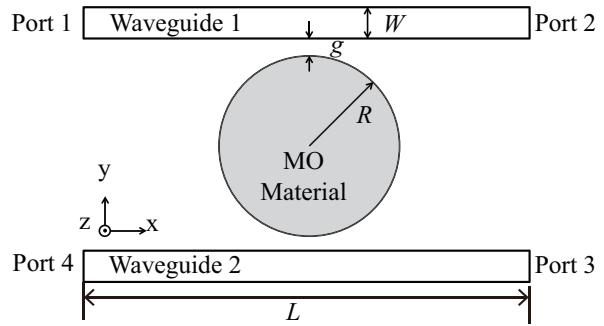}
\caption{\label{fig:epsart} The schematic of the circulator. The
circulator consists of a circular cavity with $R = 250$ um and two
straight dielectric waveguides. The straight waveguides are
symmetric about the center of the cavity. The width of the straight
waveguides is $W = 40$ um, the length is $L = 8R$ and the dielectric
constant of the waveguides is $\varepsilon _d = 5$. In addition, the
gap between the cavity and the waveguide is $g = 20$ um.}
\end{figure}

In this section, we study the interaction between the one-way SMP
cavity and the conventional waveguides, which is fundamental for
designing nonreciprocal devices. We use the one-way SMP cavity to
construct a circulator as an example. The circulator consists of a
circular cavity and two straight dielectric waveguides as shown in
Fig. 2. The parameters of the cavity are the same as those in the
third case discussed above. The straight waveguides are symmetric
about the center of the cavity. We denote the width, length and
relative permittivity of the waveguides by $W$, $L$ and $\varepsilon
_d$. The gap between the cavity and the waveguide is $g$. The
dispersion relation for the fundamental guiding TM mode
$({H_z},{E_x},{E_y})$ in the waveguide can be obtained by solving

\begin{equation}
{k_2}{\rm{ = }}\frac{{{k_1}}}{{{\varepsilon _d}}}\tan
(\frac{{{k_1}W}}{2})
\end{equation}
where ${k_1}{\rm{ = }}\sqrt {{\omega ^2}{\mu _0}{\varepsilon
_0}{\varepsilon _d} - {\beta ^2}} $, ${k_2}{\rm{ = }}\sqrt {{\beta
^2}{\rm{ - }}{\omega ^2}{\mu _0}{\varepsilon _0}}$ and $\beta$ is
the propagation constant \cite{24}. To achieve a strong interaction
between the guiding modes and the cavity mode, the phase matching
condition \cite{25} is required:
\begin{equation}
\beta ({f_m}) \cong {\beta _e}(m)(1 - \frac{g}{{2R}})
\end{equation}
where $f_m$ is the eigenfrequency of the cavity mode $\left| m
\right\rangle$, and ${\beta _e}(m)$ is the equivalent propagation
constant of the cavity mode $\left| m \right\rangle$. When the modal
field is confined well at the boundary of the cavity, ${\beta
_e}{\rm{ = }}m/R$. The term $(1 - g/2R)$ includes the effect of the
curvature at the cavity boundary on the propagation constant as seen
by the straight waveguide approximately \cite{24}. Due to the
asymmetric dispersion relation of the cavity mode, coupling occurs
in a uni-directional manner between the guiding mode and the cavity
mode within the one-way frequency ranges. We set $m = 17$ and find
that the structure with $W=40$ $\mu$m, ${\varepsilon _d}{\rm{ = }}5$
and $g=20$ $\mu$m satisfies Eq. (8). In addition, the length is
$L=8R=2000$ $\mu$m and Ports 1-4 are labeled at the end of the
waveguides.

\begin{figure}
\includegraphics[scale=0.45]{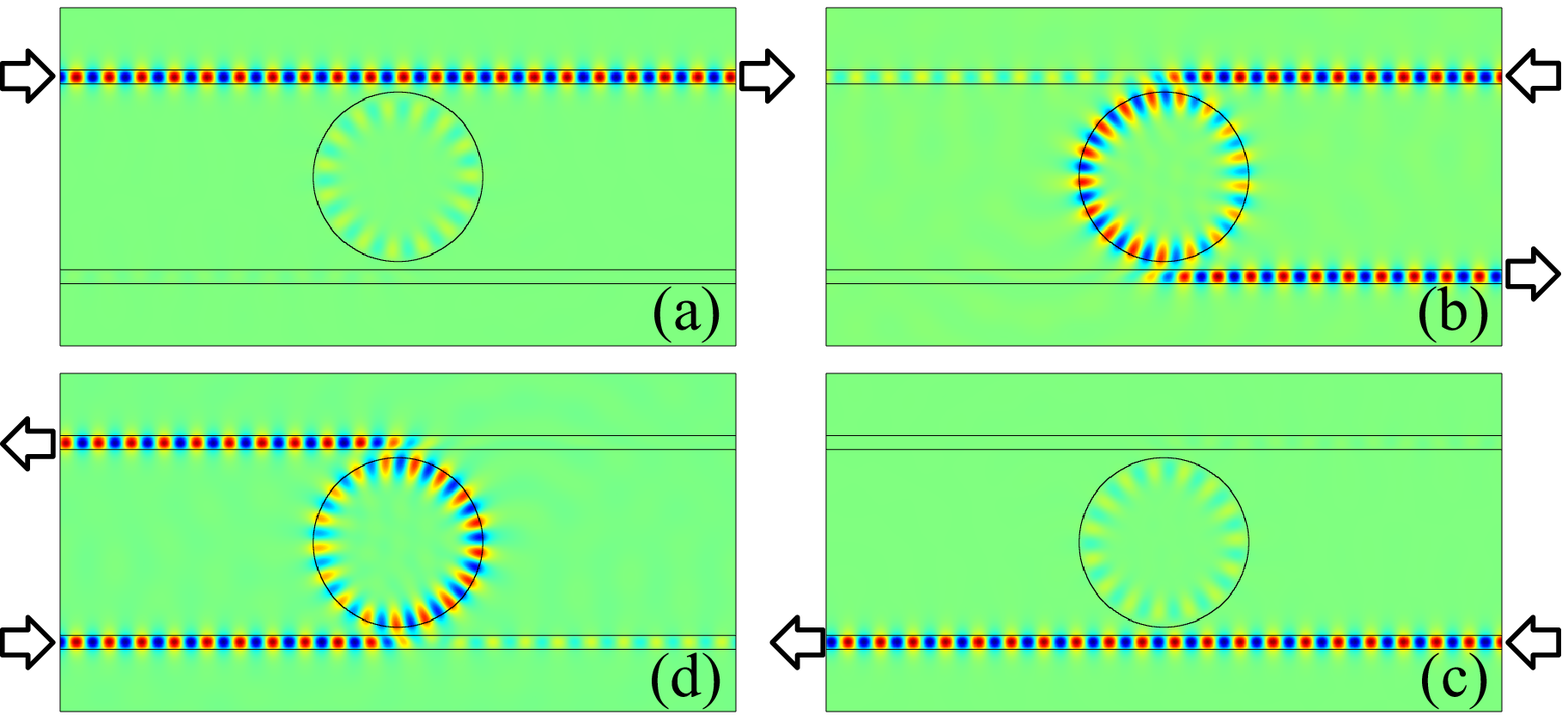}
\caption{\label{fig:epsart}Simulation result of wave propagation at
the resonance frequency for $m=+17$. In (a)-(d), input power is at
Port 1, 2, 3 and 4, respectively.}
\end{figure}

\begin{figure}
\includegraphics[scale=0.45]{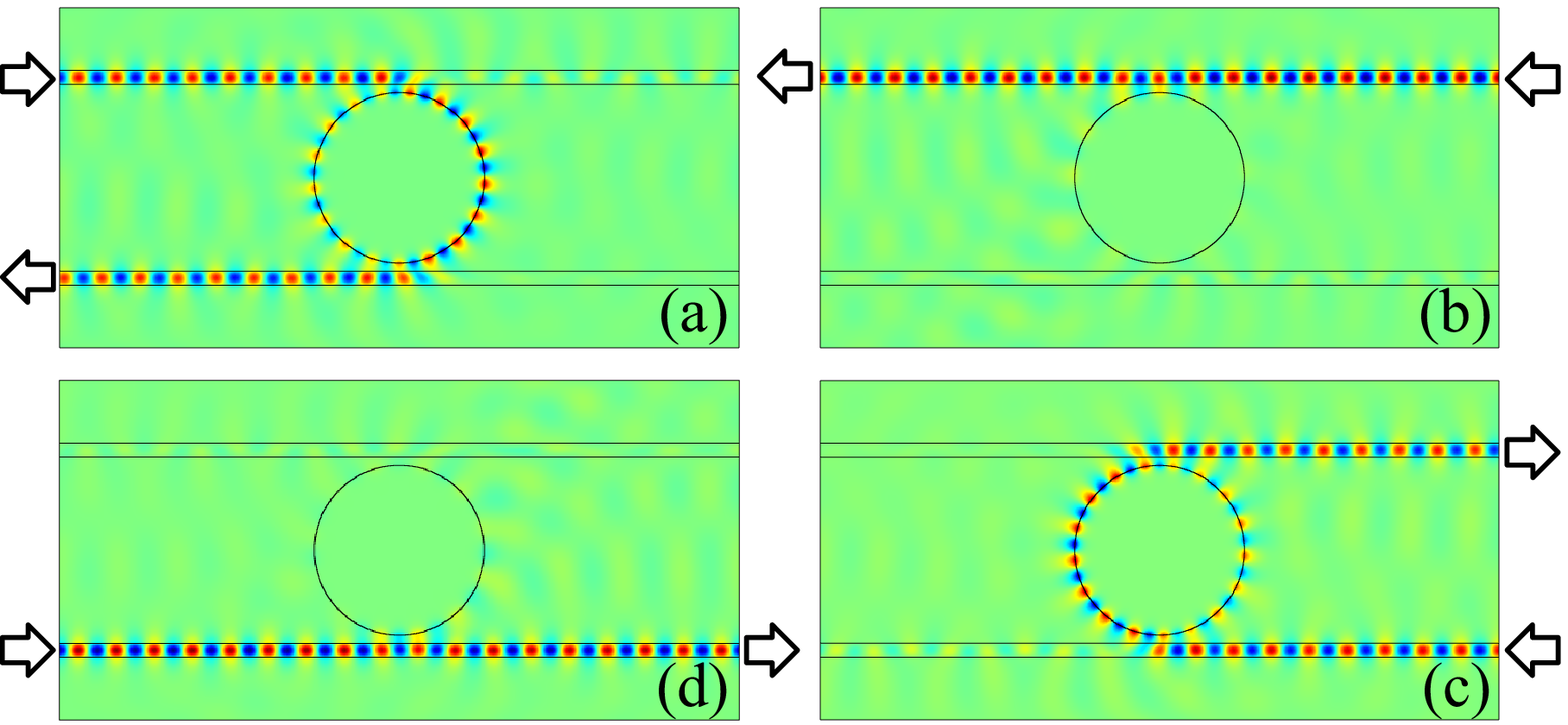}
\caption{\label{fig:epsart}Simulation result of wave propagation at
the resonance frequency for $m=-17$. In (a)-(d), input power is at
Port 1, 2, 3 and 4, respectively.}
\end{figure}

Next, we study the performance of the circulator at both one-way
frequency ranges. As the radius of the cavity is in the order of
wavelength, and the structure of the circulator is compact, we use
the FEM method to simulate the performance of the circulator. For
the upper one-way frequency range, in which only $\left| { + m}
\right\rangle $ exists, Fig. 3(a)-3(d) show the wave propagation at
the resonance frequency for $m=+17$, when the input power is at Port
1, 2, 3 and 4, respectively. The input wave from Port 1 is not
coupled with the cavity mode and goes out at Port 2 [Fig. 3(a)],
while the input wave from Port 2 is coupled with the cavity mode and
goes out at Port 3 [Fig. 3(b)]. The wave travels in the order
Port1$\to$Port2$\to$Port3$\to$Port4$\to$Port1. For the lower one-way
frequency range, in which only $\left| { - m} \right\rangle $
exists, Fig. 4(a)-4(d) show the wave propagation at the resonance
frequency for $m=-17$, when the input power is at Port1, 2, 3 and 4,
respectively. The wave travels in the order
Port1$\to$Port4$\to$Port3$\to$Port2$\to$Port1.

\begin{figure}
\includegraphics[scale=0.53]{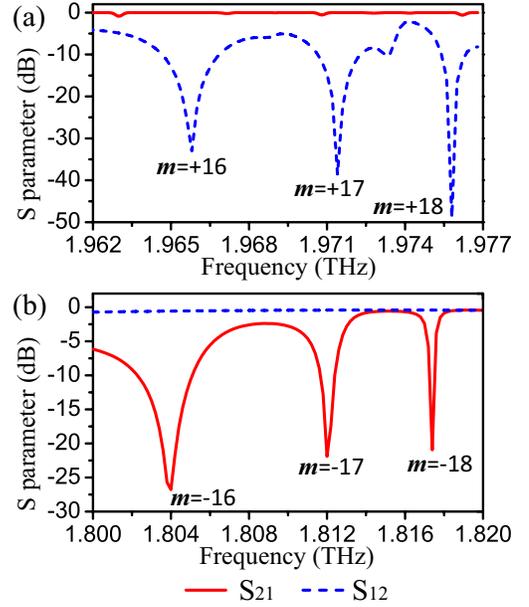}
\caption{\label{fig:epsart}Calculated S parameters for the
circulator. (a) S parameters within the upper one-way frequency
range. The dips are the resonance frequencies corresponding to $m =
+16, +17, +18$. (b) S parameters within the lower one-way frequency
range. The dips are the resonance frequencies corresponding to $m =
-16, -17, -18$. }
\end{figure}

The properties of transmission and isolation are shown by the
calculated $S$ parameters in Fig. 5. Within the upper one-way
frequency range [Fig. 5(a)], the transmission coefficient ${S_{21}}$
(red solid line) from Port 1 to Port 2 is close to 0 dB, which
indicates that the guided wave from Port 1 is not coupled with SMPs
in the cavity. The small losses (small dips of ${S_{21}}$) are
caused by a weak coupling with some other whispering-gallery-like
modes inside the cavity. This coupling can be eliminated by
increasing the gap between the cavity and the waveguides (not shown
here). The three dips of the isolation ${S_{12}}$ (blue dashed line)
are the resonance frequencies corresponding to $m =+16, +17, +18$.
The resonance frequencies are very close to the eigenfrequencies of
the cavity modes $\left| { + 16} \right\rangle$, $\left| { + 17}
\right\rangle $, $\left| { + 18} \right\rangle$, as the cavity modes
are perturbed slightly by the 2 straight waveguides. Within the
lower one-way frequency range [Fig. 5(b)], the transmission
coefficient ${S_{12}}$ (blue dashed line) from Port 2 to Port 1 is
close to 0 dB. The three dips of the isolation ${S_{21}}$ (red solid
line) are the resonance frequencies corresponding to  $m =-16, -17,
-18$. The resonance frequencies are shifted a little from their
corresponding eigenfrequencies of $\left| { - 16} \right\rangle$,
$\left| { - 17} \right\rangle $ and $\left| { - 18} \right\rangle$,
since the geometric structure is designed to fulfill the phase
matching condition Eq. (8) at $m=+17$. All the dips are lower than
-20 dB, indicating good isolation performance.

In conclusion, we have theoretically analyzed the SMP mode in a
compact circular MO cavity under a magnetic field. In such a cavity,
we have found that the different asymptotic and different cut-off
frequencies for clockwise and anti-clockwise modes lead to two
one-way frequency ranges for SMPs in the cavity. These multiple
mechanisms for achieving one-way SMPs offer more freedom, both in
the one-way direction and in the frequency range, for designing
nonreciprocal photonic components. We also have studied the
application of this cavity in a four-port circulator as an example
to show the interaction between the one-way SMP cavity and
waveguides, which is the heart of nonreciprocal devices. We believe
that the mechanisms we found in the cavity will provide people more
ways to manipulate SPs and our idea may be useful in a variety of
potential applications ranging from THz signal isolation to
isolators on chips.

This work is supported by Swedish VR grant (No. 621-2011-4620) and
AOARD. The partial support of the National Natural Science
Foundation of China (Nos. 61178062 and 91233208) is also
acknowledged. Kexin Liu thanks the China Scholarship Council (CSC)
No. 201406320056. Amir Torki thanks the Swedish Institute
Scholarship (SI).

\end{document}